\documentstyle[12pt]{elsart}
\begin{document}
\newcommand{\eq}{\begin{eqnarray}}
\newcommand{\en}{\end{eqnarray}}

\begin{frontmatter}

\date{18 September 2000}
\title{
$\pi^-p$ atom in ChPT:
strong energy-level shift}
 
\author[Tuebingen,Tomsk]{V.E. Lyubovitskij} and
\author[Bern,Tbilisi]{A. Rusetsky}
\address[Tuebingen]{Institute of Theoretical Physics, University of 
T\"{u}bingen, Auf der Morgenstelle 14, D-72076 T\"{u}bingen, Germany}
\address[Tomsk]{Department of Physics, Tomsk State University, 
634050 Tomsk, Russia}
\address[Bern]{Institute for Theoretical Physics, University of Bern,
Sidlerstrasse 5, CH-3012, Bern, Switzerland}
\address[Tbilisi]{HEPI, Tbilisi State University, 380086 Tbilisi, Georgia}

\date{\today}

\maketitle

\begin{abstract}

The general formula of the $\pi^-p$ atom strong energy-level shift
in the $1s$ state is derived in the next-to-leading 
order in the isospin breaking,
and in all orders in chiral expansion. Isospin-breaking corrections
to the level shift are explicitly evaluated at order $p^2$ in ChPT.
The results clearly demonstrate the necessity to critically reaccess
the values of the $\pi N$ scattering lengths, extracted from the energy-level
shift measurement by means of the potential model-based theoretical
analysis.

\noindent {\it PACS:} 03.65.Ge, 03.65.Nk, 11.10.St, 12.39.Fe, 13.40.Ks

\noindent {\it Keywords:} Hadronic atoms, Non-relativistic effective 
Lagrangians, Chiral Perturbation Theory

\end{abstract}

\end{frontmatter}

In the experiment performed at PSI~\cite{PSI1,PSI2}, isospin-symmetric 
$S$-wave $\pi N$ scattering lengths $a_{0+}^+$ and $a_{0+}^-$ are 
obtained\footnote{
In Refs.~\cite{PSI1,PSI2}, the quantities $b_0=a_{0+}^+$ and $b_1=-a_{0+}^-$
are used. Below, we follow the conventions of~\cite{Hoehler}} 
from the measurements of the $\pi^-p$ atom strong energy-level
shift $\varepsilon_{1s}(\pi^-p)$, 
its decay width into $\pi^0n$ state $\Gamma_{1s}(\pi^-p)$, 
and the strong energy-level shift of the $\pi^-d$ atom 
$\varepsilon_{1s}(\pi^-d)$. These three measurements are claimed to be fairly
consistent within the error bars, and - by using the technique described in 
Ref.~\cite{PSI1} - yield the strong $\pi N$ scattering lengths in an 
accuracy which is unique for hadron physics: 
$a_{0+}^+=(1.6\pm 1.3)\times 10^{-3}M_{\pi^+}^{-1}$ and
$a_{0+}^-=(86.8\pm 1.4)\times 10^{-3} M_{\pi^+}^{-1}$. 
The scattering length $a_{0+}^-$  may be  used as an input
in the Goldberger-Miyazawa-Oehme~\cite{GMO} 
sum rule to determine the $\pi NN$ coupling constant.
Recently, a new experiment has been approved 
on the pionic hydrogen~\cite{Gotta} that will allow one to measure
$\Gamma_{1s}(\pi^-p)$ with a much higher accuracy and thus, in
principle, to determine
the $\pi N$ scattering lengths from the data on pionic hydrogen alone.
This, in its turn, will enable one to vastly reduce the model-dependent 
uncertainties that come from the analysis of three-body problem~\cite{Thomas}.
Further, at CERN, the DIRAC collaboration conducts the experiment
aimed at the measurement of the $\pi^+\pi^-$ atom lifetime within the
$10~\%$ accuracy that will allow one to determine the difference of $\pi\pi$
scattering lengths $a_0-a_2$ with a $5~\%$ precision~\cite{DIRAC}. The DIRAC
measurement provides a critical test for large/small condensate scenario in 
QCD: should it turn out that the quantity $a_0-a_2$ is different from the 
value predicted in standard Chiral Perturbation Theory
(ChPT)~\cite{ChPT}, 
one has to conclude that the 
chiral symmetry breaking in QCD proceeds differently from the standard 
picture~\cite{Stern}. Finally, the DEAR collaboration~\cite{DEAR} 
at the DA$\Phi$NE facility  (Frascati) plans to measure  
the energy level shift and lifetime of the $1s$ state in $K^{-}p$ 
and $K^-d$ atoms - with considerably higher precision 
than in the previous experiment carried out at KEK~\cite{KEK} for
$K^-p$ atoms. It is expected~\cite{DEAR}
that this will result in a precise determination of the $I=0,1$ $S$-wave
scattering lengths - although, of course, one will  again be faced 
with the three-body problem already mentioned. It will be a challenge for 
theorists to extract from this new information on the $\bar{K}N$ 
amplitude at threshold a more precise value of e.g. the 
kaon-nucleon sigma term and of the strangeness content of the 
nucleon~\cite{Gensini}.

\vspace*{0.1cm}

The relations between the characteristics of hadronic atoms and
hadronic scattering lengths in the leading order in isospin breaking are given
by the well-known formulae by Deser {\it et al.}~\cite{Deser} - 
consequently, the measurement of these characteristics provides one
with the unique possibility to directly extract these scattering lengths from
the data. In order to perform the above-mentioned elaborate tests of 
QCD at low energy, the precision of the leading-order formula is,
however, not sufficient. The theoretical calculations of the hadronic atom
observables should be necessarily carried out at the accuracy that 
matches the accuracy of the experimental measurements. 
To this end, one may invoke the potential model, assuming, e.g.
that the strong interactions which are described by energy-independent
local potentials, do not violate the isospin symmetry.
This framework, applied to the case of the $\pi^+\pi^-$ atom
decay~\cite{Rasche}, failed, however, to agree with the magnitude and 
even the sign of the correction term to the leading-order formula, 
evaluated independently in ChPT~\cite{Sazdjian,Atom,Bern1,Bern2}.
The reason for this discrepancy is now well understood: the simple
potential model described above, does not include the full content of
the isospin-breaking effects in QCD. For this reason,
in our opinion, it is legitimate to question the applicability
of the potential model to the calculation
of the energy spectrum and decays of other hadronic atoms
(e.g., the $\pi^-p$ atom), as well. These calculations
should be also reexamined on the basis of low-energy effective 
theories where the isospin-breaking effects
are completely taken into account.

The present paper is aimed at establishing the precise relation
between the strong energy-level shift of the $\pi^-p$ atom in the $1s$
state, and strong $\pi N$ scattering lengths.
This relation, valid at next-to-leading order in isospin breaking,
can be used to extract $\pi N$ scattering lengths from the measured
energy spectrum of the $\pi^- p$ atom. So far, only potential model
calculations have been utilized for this purpose~\cite{PSI1}.
In the present paper, however, we use the non-relativistic effective
Lagrangian approach and incorporate all isospin-breaking effects.
The decay of $\pi^-p$ atom will be considered in a separate
publication. Note that the same approach has been
already applied to study the $\pi^+\pi^-$ atom decay in
ChPT~\cite{Bern1,Bern2}.

Our strategy is the follows. First, we write down the general
non-relativistic Lagrangian that describes the interactions between
nucleons, pions and photons at a very small momenta. 
This Lagrangian is then used - by means of Feshbach's
formalism~\cite{Feshbach} - to calculate the ground-state energy
of the $\pi^-p$ atom. The energy is expressed in terms of the
couplings of the non-relativistic Lagrangian which, in its turn, 
through the matching condition are related to the relativistic
$\pi^-p$ elastic scattering amplitude at threshold, calculated 
in ChPT. At the last step, we analyze
the isospin-symmetry breaking contributions to this amplitude at
$O(p^2)$ in ChPT that, using the already established relation
between the amplitude and the bound-state energy, provides
us with the isospin-breaking correction to the $1s$ energy level
of the $\pi^-p$ atom.

The preliminary remarks are in order.
The isospin breaking it the $\pi N$ system is due to two physically distinct
sources: electromagnetic effects being proportional to the fine structure
constant $\alpha$, and up and down quark mass difference $m_d-m_u$. 
It is convenient to introduce the common counting for these two effects.
In particular, one may define\footnote{Note
that in Refs.~\cite{Bern1,Bern2}, a slightly different convention
$\alpha\sim (m_d-m_u)^2\sim\delta$ was adopted. The counting rule is purely a
matter of convention: if we assume the same rule as in the present paper,
the $O((m_d-m_u)^2)$ contributions which emerge in 
Refs.~\cite{Bern1,Bern2}, will be
merely relegated to higher orders in the perturbation expansion.}
 the small parameter $\delta$ so that
$\alpha\sim (m_d-m_u)\sim\delta$. With this definition, both 
$\Delta M_\pi^2=O(\delta)$ and $m_p-m_n=O(\delta)$.
Further, in the leading order in the fine-structure constant, the 
binding energy of the $\pi^- p$ atom ground state is given by the 
well-known non-relativistic Coulomb-Schr\"{o}dinger value 
$E_1=-\frac{1}{2}\mu_c\alpha^2$, where $\mu_c$ denotes the reduced mass 
of $\pi^- p$ system. The leading-order strong energy-level shift is given 
by the strong amplitude times the square of the Coulomb wave function at 
the origin and, consequently, is of order $O(\alpha^3)$. We are 
interested in the leading order isospin-breaking corrections to the 
strong shift - thus we have to evaluate the bound-state energy at 
$O(\delta^4)$ that stands either for $O(\alpha^4)$, or for 
$O(\alpha^3(m_d-m_u))$\footnote{There is one exception from this counting 
rule. We include the effect due to the electron vacuum polarization as
well. Formally this contribution is of 
order $\alpha^5$, but it is amplified by the large factor $(M_{\pi^+}/m_e)^2$.}.

We emphasize that in this paper we deal with the shift of the
ground-state level while the quantity that is experimentally measured is the 
transition energy between $3p-1s$ levels. The strong shift in the $3p$ state
is, however, very small (see~\cite{PSI1} and references
therein), and for the time being is neglected in our approach. 

Now, we start with writing down the non-relativistic Lagrangian which
will be used to calculate the energy of the $\pi^- p$ atom.
This Lagrangian, in general, consists of an infinite tower of
operators with an increasing mass dimension - it contains all possible
terms that are allowed by discrete $P,~C,~T$ symmetries, rotational 
invariance and gauge
invariance. The convenient building blocks for this Lagrangian,
therefore, are: the non-relativistic pion and nucleon fields,
their covariant derivatives, and electric (${\bf E}$) and magnetic
(${\bf B}$) fields. For the general procedure of the derivation, 
we refer to Ref.~\cite{Kinoshita}
which deals with the same problem in detail, in the context of QED, and merely
display the result   

\eq\label{Lagr-ini}
{\cal L}&=&-\frac{1}{4}\, F_{\mu\nu}F^{\mu\nu}+
\psi^\dagger\,\biggl\{i{\cal D}_t-m_p+\frac{{\cal D}^2}{2m_p}
+\frac{{\cal D}^4}{8m_p^3}+\cdots
\nonumber\\[2mm]
&-&c_p^F\,\frac{e\sigma{\bf B}}{2m_p}
-c_p^D\,\frac{e({\cal D}{\bf E}-{\bf E}{\cal D})}{8m_p^2}
-c_p^S\,\frac{ie\sigma({\cal D}\times{\bf E}-{\bf E}\times{\cal D})}{8m_p^2}
+\cdots\biggr\}\,\psi
\nonumber\\[2mm]
&+&\chi^\dagger\,\biggl\{i\partial_t-m_n+\frac{\nabla^2}{2m_n}
+\frac{\nabla^4}{8m_n^3}+\cdots
\nonumber\\[2mm]
&-&c_n^F\,\frac{e\sigma{\bf B}}{2m_n}
-c_n^D\,\frac{e(\nabla{\bf E}-{\bf E}\nabla)}{8m_n^2}
-c_n^S\,\frac{ie\sigma(\nabla\times{\bf E}-{\bf E}\times\nabla)}{8m_n^2}
+\cdots\biggr\}\,\chi
\nonumber\\[2mm]
&+&\sum_\pm\pi_\pm^\dagger\,\biggl\{iD_t-M_{\pi^+}
+\frac{{\bf D}^2}{2M_{\pi^+}}+\frac{{\bf D}^4}{8M_{\pi^+}^3}+\cdots
\nonumber\\[2mm]
&\mp& c^R\,\frac{e({\bf D}{\bf E}-{\bf E}{\bf D})}{6M_{\pi^+}^2}
+\cdots\biggr\}\,\pi_\pm 
\nonumber\\[2mm]
&+&\pi_0^\dagger\,\biggl\{i\partial_t-M_{\pi^0}+\frac{\nabla^2}{2M_{\pi^0}}
+\frac{\nabla^4}{8M_{\pi^0}^3}+\cdots\biggr\}\,\pi_0
\nonumber\\[2mm]
&+&d_1\,\psi^\dagger\psi\,\pi_-^\dagger\pi_-+
d_2(\psi^\dagger\chi\,\pi_-^\dagger\pi_0+h.c.)
+d_3\chi^\dagger\chi\,\pi_0^\dagger\pi_0+\cdots\, .
\en
Here $F_{\mu\nu}$ stands for the electromagnetic field strength tensor,
$\psi$, $\chi$, $\pi_{\pm}$ and $\pi_0$ denote the non-relativistic field
operators for proton, neutron, charged and neutral pion fields,
and ${\cal D}_t\, \psi=\partial_t\, \psi-ieA_0\, \psi$,   
${\cal D}\,\psi= \nabla\, \psi+ie{\bf A}\, \psi$,
$D_t\, \pi_\pm=\partial_t\,\pi_\pm\mp ieA_0\,\pi_\pm$,
${\bf D}\, \pi_\pm= \nabla\,\pi_\pm\pm ie{\bf A}\,\pi_\pm$ are the covariant
derivatives acting on the proton and charged pion fields, respectively.
The ellipses stand for the higher-dimensional operators and the counterterms
that are needed to render the off-shell Green functions finite in the
perturbation theory.
The couplings
$c_p^F,~c_p^D,~c_p^S,~c_n^F,~c_n^D,~c_n^S,~c^R,~d_1,~d_2,~d_3,\cdots$
should be determined from matching of non-relativistic and relativistic
amplitudes (see below). In general, the matching procedure determines the 
same combinations of the non-relativistic couplings that are needed in the
bound-state calculations~\cite{Lamb}.

The Lagrangian displayed in Eq.~(\ref{Lagr-ini}), contains an infinite tower
of operators with the increasing mass dimension. This Lagrangian is used in
the perturbation theory to calculate the scattering amplitudes in the
one-nucleon, one-pion and one-nucleon-one-pion sectors, with any number of
photons, to a given accuracy in the coupling constant $e$ and in the 
expansion in external 3-momenta.
At any order in the perturbative expansion, these amplitudes
can be renormalized by adding the appropriate counterterms to the
Lagrangian~(\ref{Lagr-ini}). 
Moreover, these amplitudes coincide with the same
amplitudes calculated in the relativistic theory, provided the couplings
in the non-relativistic Lagrangian are matched in a sufficient accuracy
in $e$. The perturbative calculations in the non-relativistic theory are
carried out using the ``threshold expansion''~\cite{Beneke} in all Feynman
integrals - this ensures that the counting rules are obeyed so that the 
higher-dimensional operators do not contribute to the momentum expansion 
of the amplitudes at lower orders. In particular, this ensures that the mass
parameters in the Lagrangian~(\ref{Lagr-ini}) coincide with the physical
masses of the particles - there is no mass renormalization.
Further, the same non-relativistic Lagrangian is
used to calculate the bound-state characteristics. Here, the powers of
external momenta ``translate'' into the powers of $e$ and the threshold
expansion should be again used to ensure the validity of counting rules.
For the details, we refer the reader to Ref.~\cite{Lamb} and references
therein. In actual calculations, one always deals with the truncated
Lagrangian. In particular,
in the present paper we aim at the evaluation of the ground-state energy
of the $\pi^-p$ atom at the accuracy $O(\delta^4)$. It can be demonstrated
that then it is necessary to construct the non-relativistic 
Lagrangian that describes correctly all contributions to the $\pi^-p$ 
elastic scattering amplitude at $O(\delta)$, which do not vanish at
physical threshold. The Lagrangian displayed in Eq.~(\ref{Lagr-ini}), is
sufficient for this purpose. It can be shown that all operators that can be
constructed (including the counterterms), 
except the ones given in Eq.~(\ref{Lagr-ini}), will not
contribute to the $\pi^-p$ scattering amplitude at $O(\delta)$ at threshold
neither at tree level, nor through loops.  

\vspace*{1.1cm}

Below, we consider briefly the non-relativistic couplings
$c_p^F$, $c_p^D$, $c_p^S$, $c_n^F$, $c_n^D$, $c_n^S$, $c^R$.
At the accuracy we are working,
it is sufficient to match these constants at $O(e^0)$. 
To this end, one may consider the nucleon and pion electromagnetic
form-factors in the external field $A_\mu$. The necessary details, including
different normalization of states in the relativistic and 
non-relativistic theories, are provided in~\cite{Kinoshita}, where the same
problem is treated in the context of QED. Below, we merely display the result
of the matching
\eq\label{matching-c_i}
c_p^F=1+\mu_p\hspace*{3.45cm}&,&\quad
c_n^F=\mu_n\nonumber\\[2mm]
c_p^D=1+2\mu_p+\frac{4}{3}\,\, m_p^2<r_p^2>\quad&,&\quad
c_n^D=2\mu_n+\frac{4}{3}\,\, m_n^2<r_n^2>\nonumber\\[2mm]
c_p^S=1+2\mu_p\hspace*{3.25cm}&,&\quad
c_n^S=2\mu_n\nonumber\\[2mm]
c^R=M_{\pi^+}^2<r_\pi^2>\hspace*{2.4cm}&,&
\en
where $\mu_p$ and $\mu_n$ denote the anomalous magnetic momenta of proton and
neutron, respectively, and $<r^2_p>,~<r^2_n>,~<r^2_\pi>$ stand for the
charge radii of proton, neutron and charged pion.
The remaining constants $d_i,~i=1\cdots 3$ are determined from matching the 
$\pi N$ scattering amplitudes in different channels.

The Lagrangian~(\ref{Lagr-ini}) still contains the terms that do not
contribute to the ground-state energy level shift of the $\pi^-p$ atom
in the accuracy we are working. In order to simplify this Lagrangian,
in the Coulomb gauge 
we exclude the $A_0$ field by using the equations of motion, and
neglect high-dimensional operators that arise in a result of this
operation. Further, we neglect the operators that contribute to the
$\pi N$ scattering sectors with non-zero total charge, and to the
spin-flip part of the $\pi^-p$ elastic scattering amplitude (since the
ground state of the $\pi^-p$ atom is an $S$-wave state, the spin-flip
part does not contribute due to the rotational invariance). In a
result, we arrive at the simplified Lagrangian which is better suited
for the bound-state calculations 
\eq\label{Lagr-fin}
{\cal L}'&=&\frac{1}{2}\,\dot{\bf A}^2-\frac{1}{2}\, {\bf B}^2
-e^2\,(\psi^\dagger\psi)\,\triangle^{-1}\,(\pi_-^\dagger\pi_-)
\nonumber\\[2mm]
&+&\psi^\dagger\,\biggl\{i\partial_t-m_p+\frac{\nabla^2}{2m_p}
+\frac{\nabla^4}{8m_p^3}+\frac{ie}{2m_p}\,(\nabla{\bf A}+{\bf A}\nabla)
\biggr\}\psi
\nonumber\\[2mm]
&+&\chi^\dagger\,\biggl\{i\partial_t-m_n+\frac{\nabla^2}{2m_n}
+\frac{\nabla^4}{8m_n^3}\biggr\}\chi
\nonumber\\[2mm]
&+&\sum_\pm\pi_\pm^\dagger\,\biggl\{i\partial_t-M_{\pi^+}
+\frac{\nabla^2}{2M_{\pi^+}}+\frac{\nabla^4}{8M_{\pi^+}^3}
\pm\frac{ie}{2M_{\pi^+}}\,(\nabla{\bf A}+{\bf A}\nabla)
\biggr\}\pi_\pm
\nonumber\\[2mm]
&+&\pi_0^\dagger\,\biggl\{i\partial_t-M_{\pi^0}+\frac{\nabla^2}{2M_{\pi^0}}
+\frac{\nabla^4}{8M_{\pi^0}^3}\biggr\}\,\pi_0
\nonumber\\[2mm]
&+&d_1'\,\psi^\dagger\psi\,\pi_-^\dagger\pi_-+
d_2(\psi^\dagger\chi\,\pi_-^\dagger\pi_0+h.c.)
+d_3\chi^\dagger\chi\,\pi_0^\dagger\pi_0\, ,
\en 
where
\eq\label{lambda}
d_1'=d_1-e^2\,\lambda\, ,\quad\quad
\lambda=\biggl(\frac{c_p^D}{8m_p^2}+\frac{c^R}{6M_{\pi^+}^2}\biggr)
=\frac{1+2\mu_p}{8m_p^2}+\frac{1}{6}\,(\langle r_p^2\rangle+
\langle r_\pi^2\rangle)
\en

For the calculation of the energy-level shift of $\pi^- p$ atom ground state
we again, as in Ref.~\cite{Bern1} use the Feshbach's formalism~\cite{Feshbach}.
To this end, we first obtain the total Hamiltonian of the system from
the Lagrangian~(\ref{Lagr-fin}) by using the canonical formalism
\eq\label{Hamiltonian}
{\bf H}={\bf H}_{\rm 0}+{\bf H}_{\rm C}+{\bf H}_{\rm R}
+e{\bf H}_\gamma+{\bf H}_{\rm S}+e^2\lambda{\bf H}_\lambda
={\bf H}_{\rm 0}+{\bf H}_{\rm C}+{\bf V}\, ,
\en
where ${\bf H}_0$ is the free Hamiltonian describing photons and  
non-relativistic pions and nucleons. Further,
${\bf H}_\Gamma=\int d^3{\bf x}\, H_\Gamma,
~\Gamma={\rm C,R,}\gamma,{\rm S},\lambda$, and
\eq\label{Hamiltonian-int}
H_{\rm C}&=&
e^2\,(\psi^\dagger\psi)\,\triangle^{-1}\,(\pi_-^\dagger\pi_-)\, ,
\nonumber\\[2mm]
H_{\rm R}&=&
-\psi^\dagger\, \frac{\nabla^4}{8m_p^3} \,\psi
-\chi^\dagger\, \frac{\nabla^4}{8m_n^3} \,\chi
-\sum_\pm\pi_\pm^\dagger\, \frac{\nabla^4}{8M_{\pi^+}^3} \,\pi_\pm
-\pi_0^\dagger\, \frac{\nabla^4}{8M_{\pi^0}^3} \,\pi_0\, ,
\nonumber\\[2mm]
H_\gamma&=&
-\frac{i}{2m_p}\,\psi^\dagger\, (\nabla{\bf A}+{\bf A}\nabla)\, \psi
-\sum_\pm\frac{\pm i}{2M_{\pi^+}}\,
\pi_\pm^\dagger\, (\nabla{\bf A}+{\bf A}\nabla)\, \pi_\pm\, ,
\nonumber\\[2mm]
H_{\rm S}&=&
-d_1\,\psi^\dagger\psi\,\pi_-^\dagger\pi_-
-d_2(\psi^\dagger\chi\,\pi_-^\dagger\pi_0+h.c.)
-d_3\chi^\dagger\chi\,\pi_0^\dagger\pi_0\, ,
\nonumber\\[2mm]
H_\lambda&=&\psi^\dagger\psi\,\pi_-^\dagger\pi_-\, .
\en

We treat the problem in the perturbation theory: 
${\bf H}_0+{\bf H}_{\rm C}$ is the unperturbed Hamiltonian,
whereas ${\bf V}$ is considered as a perturbation.
The ground-state solution of the unperturbed Schr\"{o}dinger equation
in the CM frame\\
$(\tilde E_1-{\bf H}_0-{\bf H}_{\rm C})|\Psi_0({\bf 0};s)\rangle=0$, with 
$\tilde E_1=m_p+M_{\pi^+}+E_1$, is given by
\eq\label{Psi_0}
|\Psi_0({\bf 0};s)\rangle=\int\frac{d^3{\bf p}}{(2\pi)^3}\,\,\Psi_0({\bf p})
\,\, b_+^\dagger({\bf p},s)\,a_-^\dagger(-{\bf p})\, |0\rangle\, ,
\en
where $a_-^\dagger({\bf p})$ and $b_+^\dagger({\bf p},s)$ 
denote creation operators for non-relativistic $\pi^-$ and proton
acting on the Fock space vacuum, 
$s$ is the projection of the proton (atom) spin, and 
$\Psi_0({\bf p})$ stands for the non-relativistic Coulomb wave function in the
momentum space.

Now, we are going to evaluate the energy-level shift of the ground
state due to the perturbation Hamiltonian ${\bf V}$.
To this end, we again, as in Ref.~\cite{Bern1}, look for the poles
of the scattering operator ${\bf T}(z)=({\bf H}_{\rm C}+{\bf V})+ 
({\bf H}_{\rm C}+{\bf V}){\bf G}_0(z){\bf T}(z)$ on the second 
Riemann sheet in the complex $z$-plane: the real and imaginary parts of the
pole position give, by definition, the energy and the decay width of the
metastable bound state. The free and Coulomb Green operators are defined as
${\bf G}_0(z)=(z-{\bf H}_0)^{-1}$ and 
${\bf G}(z)=(z-{\bf H}_0-{\bf H}_{\rm C})^{-1}$, respectively.
Further, we define the ``Coulomb-pole removed'' Green function as
$\hat {\bf G}(z)={\bf G}(z)(1-{\bf \Pi})$, where ${\bf \Pi}$ denotes the
projector onto the Coulomb ground state $\Psi_0$~(\ref{Psi_0}).
The $\pi N$ scattering states in the sector with the total charge $0$ are
defined as
$|{\bf P},{\bf p};s\rangle_+=b_+^\dagger({\bf p}_1,s)\,
a_-^\dagger({\bf p}_2)\, |0\rangle$ and
$|{\bf P},{\bf p};s\rangle_0=b_0^\dagger({\bf p}_1,s)\,
a_0^\dagger({\bf p}_2)\, |0\rangle$
($a_0^\dagger({\bf p})$ and $b_0^\dagger({\bf p},s)$ denote the creation
operators for the $\pi^0$ meson and neutron, respectively).
The center-of mass and relative momenta are defined by
${\bf P}={\bf p}_1+{\bf p}_2$,
${\bf p}=\eta^{(A)}_2{\bf p}_1-\eta^{(A)}_1{\bf p}_2$,
$A=+,0$, and 
$\eta^{(+)}_1=m_p/(m_p+M_{\pi^+})$, $\eta^{(+)}_2=M_{\pi^+}/(m_p+M_{\pi^+})$,
$\eta^{(0)}_1=m_n/(m_n+M_{\pi^0})$, $\eta^{(0)}_2=M_{\pi^0}/(m_n+M_{\pi^0})$.
We remove the CM momenta from the matrix elements of any operator 
${\bf R}(z)$ by introducing the notation
\eq\label{CM}
_A\langle {\bf P},{\bf q};s|{\bf R}(z)|
{\bf 0},{\bf p};s'\rangle_B=
(2\pi)^3\delta^3({\bf P})\,
({\bf q},s|{\bf r}_{AB}(z)|{\bf p};s')\, .
\en

In general, the expression for the matrix elements of ${\bf r}_{AB}(z)$ 
contains the spin-nonflip ad the spin-flip parts that are defined in the
following manner
\eq\label{flip-nonflip}
({\bf q};s|{\bf r}_{AB}(z)|{\bf p};s')=
\delta_{ss'}({\bf q}|{\bf r}^{\bf n}_{AB}(z)|{\bf p})
+i(\sigma_{ss'}\cdot [\,{\bf p}\times {\bf q}\, ])\, 
({\bf q}|{\bf r}^{\bf f}_{AB}(z)|{\bf p})\, .
\en

Below, we define the ``Coulomb-pole removed'' transition operator
that satisfies the equation
\eq\label{pole-removed}
{\bf M}(z)={\bf V}+{\bf V}\hat {\bf G}(z){\bf M}(z)\, .
\en

With the use of the Feshbach's formalism and the definitions given above, 
it is straightforward to 
demonstrate that the scattering operator ${\bf T}(z)$ develops the pole at
$z=\bar z$ where $\bar z$ is the solution of the following equation
\eq\label{pole-position}
\bar z-\tilde E_1-(\Psi_0|{\bf m}^{\bf n}_{++}(\bar z)|\Psi_0)=0\, ,
\en
where $({\bf p}|\Psi_0)=\Psi_0({\bf p})$ and ${\bf m}^{\bf n}_{++}(z)$ is
related to ${\bf M}(z)$ through the definitions~(\ref{CM}) and
(\ref{flip-nonflip}). 

In order to get the shift of the ground-state energy,
the quantity ${\bf m}^{\bf n}_{++}(z)$ is determined perturbatively
from Eq.~(\ref{pole-removed}). The iteration series that emerge from
Eq.~(\ref{pole-removed}), can be truncated since the higher-order
terms do not contribute to the energy at $O(\delta^4)$. Using the
explicit expression of ${\bf V}$ given by Eq.~(\ref{Hamiltonian}),
replacing $\hat {\bf G}(z)$ by ${\bf G}_0(z)$ whenever possible,
and retaining only those terms that contribute at the accuracy we are
working, the operator ${\bf M}(z)$ can be written in the form
${\bf M}(z)={\bf U}(z)+{\bf W}(z)$, where
\eq\label{uw}
{\bf U}(z)&=&{\bf H}_{\rm R}
+e^2{\bf H}_\gamma {\bf G}_0(z){\bf H}_\gamma
+e^2\lambda{\bf H}_\lambda\, ,
\nonumber\\[2mm]
{\bf W}(z)&=&{\bf H}_{\rm S}+{\bf H}_{\rm S}\hat {\bf G}(z){\bf H}_{\rm S}
+{\bf H}_{\rm S}{\bf G}_0(z){\bf H}_{\rm S}{\bf G}_0(z){\bf H}_{\rm S}\, ,
\en

At the accuracy $O(\delta^4)$, the energy of the bound state is equal to
$\bar z=\tilde E_1+\Delta E^{\rm em}_1+\epsilon_{1s}$, where
\eq\label{solution-key}
\Delta E^{\rm em}_1
={\rm Re}\, (\Psi_0|{\bf u}^{\bf n}_{++}(\tilde E_1)|\Psi_0)+E^{\rm vac}\, ,
\quad
\epsilon_{1s}
={\rm Re}\, (\Psi_0|{\bf w}^{\bf n}_{++}(\tilde E_1)|\Psi_0)\, .
\en
Here ${\bf u}^{\bf n}_{++}(z)$, ${\bf w}^{\bf n}_{++}(z)$ are related to
${\bf U}(z)$, ${\bf W}(z)$ through the definitions (\ref{CM}) and
(\ref{flip-nonflip}), and
$E^{\rm vac}$ stands for the contribution due to the electron vacuum
polarization (the corresponding term is added ``by hand'', see below).

Using explicit expressions~(\ref{Hamiltonian-int}), we obtain
\eq\label{Re}
&&\hspace*{-0.6cm}
{\rm Re}\, (\Psi_0|{\bf u}^{\bf n}_{++}(\tilde E_1)|\Psi_0)=
-\frac{5}{8}\,\alpha^4\mu_c^4\,
\frac{m_p^3+M_{\pi^+}^3}{m_p^3M_{\pi^+}^3}
-\frac{\alpha^4\mu_c^3}{m_p M_{\pi^+}}+4\alpha^4\mu_c^3\lambda\, ,
\\[2mm]
&&\hspace*{-0.6cm}
{\rm Re}\, (\Psi_0|{\bf w}^{\bf n}_{++}(\tilde E_1)|\Psi_0)=
\frac{\alpha^3\mu_c^3}{\pi}\, \biggl(
-d_1+d_1^2\, (\,\xi+\frac{\alpha\mu_c^2}{\pi}\, (\ln\alpha-1))
+\frac{\mu_0^2q_0^2}{4\pi^2}\, d_2^2d_3\biggr)\, ,
\nonumber
\en
where $\mu_0$ is the reduced mass of the $\pi^0n$ system, and
\eq\label{xi}
\xi&=&\frac{\alpha\mu_c^2}{2\pi}\biggl\{
(\mu^2)^{d-3}\biggl(\frac{1}{d-3}-\Gamma'(1)-\ln 4\pi\biggr)
+\ln\frac{(2\mu_c)^2}{\mu^2}-1\biggr\}\, ,
\nonumber\\[2mm]
q_0&=&\bigl[ 2\mu_0(m_p+M_{\pi^+}-m_n-M_{\pi^0})\bigr]^{1/2}
\en
Here $d$ and $\mu$ denote the dimension of space and the scale of the
dimensional regularization, respectively 
(we use the dimensional regularization scheme throughout the paper).

The vacuum polarization contribution, which is calculated separately, should 
be added to the electromagnetic energy shift. The modification of Coulomb 
potential due to electron vacuum polarization is given by the well-known  
expression~\cite{Atom}
\eq\label{vacuum-potential}
\Delta V^{\rm vac}(r)=-\frac{4}{3}\,\alpha^2
\int\frac{d^3{\bf k}}{(2\pi)^3}\,{\rm e}^{i{\bf k}{\bf r}}
\int_{4m_e^2}^\infty\frac{ds}{s+{\bf k}^2}\,
\frac{1}{s}\biggl(1+\frac{2m_e^2}{s}\biggr)
\sqrt{1-\frac{4m_e^2}{s}}\, ,
\en
where $m_e$ denotes the electron mass.
Applying the first-order perturbation theory, we find
\eq\label{vac-pol}
&&\hspace*{-0.3cm}
E^{\rm vac}=\int d^3{\bf r}\,\Psi_0^2({\bf r})\,
\Delta V^{\rm vac}(r)
=-\frac{\alpha^3\mu_c}{3\pi}\,\eta^2\Phi(\eta),\quad
\eta=\frac{\alpha\mu_c}{m_e}\, ,
\\[2mm]
&&\hspace*{-0.3cm}
\eta^2\Phi(\eta)=\frac{1}{\eta^3}\biggl(2\pi-4\eta+\frac{3\pi}{2}\,\eta^2
-\frac{11}{3}\,\eta^3\biggr)
+\frac{2\eta^4-\eta^2-4}{\eta^3\sqrt{\eta^2-1}}\,
\ln(\eta+\sqrt{\eta^2-1})\, .
\nonumber
\en

The calculation of the electromagnetic energy-level shift is now complete.
In order to compare our results with those given in Ref.~\cite{PSI1}, it is
convenient to define
\eq\label{em-comparison}
\tilde E_1+\Delta E_1^{\rm em}=E^{\rm KG}+E^{\rm fin}+E^{\rm vac}+
E^{\rm rel}\, ,
\en
where
\eq\label{em-individual}
E^{\rm KG}&=& 
-\frac{1}{2}\, \mu_c \alpha^2\,\biggl(1+\frac{5\alpha^2}{4}\biggr)\, ,
\nonumber\\[2mm]
E^{\rm fin}&=&\frac{2}{3}\,\alpha^4\mu_c^3\,
(\langle r_p^2\rangle+\langle r_\pi^2\rangle)\, ,
\\[2mm]
E^{\rm rel}&=&-\frac{1}{2}\,\mu_c\alpha^4\,\biggl(
\frac{\mu_c}{4(m_p+M_{\pi^+})}+\frac{m_p^2}{(m_p+M_{\pi^+})^2}-1
-\frac{2\mu_p M_{\pi^+}^2}{(m_p+M_{\pi^+})^2}\biggr)\, ,
\nonumber
\en
and $E^{\rm vac}$ is given by Eq.~(\ref{vac-pol}). In Table~1
we collect various contributions to the electromagnetic energy-level shift
calculated using the same input parameters, as in 
Ref.~\cite{PSI1}. As can be seen
from Table~1, the calculated values of the electromagnetic 
shift almost coincide.

\begin{figure}[t]
{\small
{\bf Table 1.}
Contributions to the electromagnetic binding energy of the $\pi^-p$
atom (eV). Higher-order corrections: vacuum 
polarization correction (order $\alpha^3$) and vertex correction have not 
been calculated.

\vspace*{.3cm}

\noindent
\begin{tabular}{ l l r r }
\hline\hline
Corrections & Notation & \hspace*{1.2cm} Ref.~\cite{PSI1} & \hspace*{1.2cm} Our     \\
\hline
Point Coulomb, KG equation              &$E^{\rm KG}$   & $-3235.156$ & $-3235.156$ \\
Finite size effect (proton, pion)       &$E^{\rm fin}$  & $0.102$     & $0.100$     \\
Vacuum polarization, order $\alpha^2$   &$E^{\rm vac}$  & $-3.246$    & $-3.241$    \\
Relativistic recoil, proton spin and    &               &             &             \\
anomalous magnetic moment               &$E^{\rm rel}$  & $0.047$     & $0.047$     \\
Vacuum polarization, order $\alpha^3$   &               & $-0.018$    &             \\
Vertex correction                       &               & $0.007$     &             \\
\hline\hline
\end{tabular}

}  
                                        
\vspace*{.5cm}

\end{figure}

In order to complete the calculation of the strong energy-level shift, one 
has to match at the accuracy $O(\delta)$ the particular combination of 
the non-relativistic couplings $d_i,~i=1\cdots 3$ that appears in 
Eq.~(\ref{Re}). To this end, we consider the scattering operator
\eq\label{TR}
&&{\bf T}_{\rm R}(z)={\bf V}_{\rm R}+{\bf V}_{\rm R}{\bf G}_{\rm R}(z)
{\bf T}_{\rm R}(z)\, ,
\nonumber\\[2mm]
&&{\bf V}_{\rm R}={\bf H}_{\rm C}+e{\bf H}_\gamma+{\bf H}_{\rm S}+
e^2\lambda{\bf H}_\lambda\, ,\quad
{\bf G}_{\rm R}(z)=(z-{\bf H}_0-{\bf H}_{\rm R})^{-1}\, .
\en

In the scattering operator ${\bf T}_{\rm R}(z)$, all kinematical insertions
contained in ${\bf H}_{\rm R}$ are summed up in the external lines
(see~\cite{Lamb} for the details). We calculate the matrix element of the
scattering operator ${\bf T}_{\rm R}(z)$ between the $\pi^-p$ states at 
$O(\delta)$. After removing the CM momentum, the spin-nonflip part of 
this matrix element on energy shell is equal to
\eq\label{T-delta}
&&\hspace*{-0.9cm}
({\bf q}|{\bf t}^{\bf n}_{{\rm R},++}(z)|{\bf p})=
-\frac{4\pi\alpha}{|{\bf q}-{\bf p}|^2}-\frac{4\pi\alpha}{4m_pM_{\pi^+}}\,
\frac{({\bf q}+{\bf p})^2}{|{\bf q}-{\bf p}|^2}
+{\rm e}^{2i\alpha\theta_{\rm C}(|{\bf p}|)}\, 
({\bf q}|\bar {\bf t}^{\bf n}_{{\rm R},++}(z)|{\bf p})\, ,
\nonumber\\
&&
\en
where the (divergent) Coulomb phase in the
dimensional regularization scheme is given by
\eq\label{phase}
\theta_{\rm C}(|{\bf p}|)=\frac{\mu_c}{|{\bf p}|}\,
\mu^{d-3}\,\biggl(\frac{1}{d-3}-\frac{1}{2}(\Gamma'(1)+\ln 4\pi)\biggr)
+\frac{\mu_c}{|{\bf p}|}\,\ln\frac{2|{\bf p}|}{\mu}\, ,
\en 
and
\eq\label{T-threshold}
{\rm Re}\,({\bf q}|\bar {\bf t}^{\bf n}_{{\rm R},++}(z)|{\bf p})=
-\frac{\pi\alpha\mu_c d_1}{|{\bf p}|}
+\frac{\alpha\mu_c^2d_1^2}{\pi}\, \ln\frac{|{\bf p}|}{\mu_c}+e^2\lambda
\nonumber\\[2mm]
-d_1+\frac{\mu_0^2 q_0^2}{4\pi^2}\,d_2^2d_3
+d_1^2\xi+\cdots\, ,
\en
where ellipses stand for the terms that vanish at threshold.

In order to carry out the matching, we consider the elastic $\pi^-p$
scattering amplitude $T_{\pi^-p\rightarrow\pi^-p}$, calculated in the
relativistic theory at $O(\delta)$\footnote{
Note that, in order to have the same sign as in the non-relativistic
definition, the sign of the relativistic amplitude is defined by $S=1-iT$.}.
This amplitude can be uniquely decomposed
into the piece containing all diagrams that are made disconnected by 
cutting one photon line, and the remainder:
$T_{\pi^-p\rightarrow\pi^-p}=T_{ex}+T_1$, with
\eq\label{splitting}
&&\hspace*{-0.8cm}
T_{ex}=\frac{e^2}{t}\,\bar u(p_1s)\biggl\{ \gamma_\mu F_1(t)+i\sigma_{\mu\nu}
(p_1-q_1)^\nu\,\frac{F_2(t)}{2m_p}\,\biggr\}\, u(q_1s')\,(p_2+q_2)^\mu\,f(t)
\, ,
\en
where $(q_1,s),~q_2$ and $(p_1,s'),~p_2$ are the four-momenta and spin of
outgoing/incoming particles, $t=(q_1-p_1)^2$, 
and  $F_i(t)$ and $f(t)$ denote, respectively, 
the nucleon and pion electromagnetic form factors.
Then, from the matching condition
$T_{\pi^-p\rightarrow\pi^-p}=(2E_q2E_p2w_q2w_p)^{1/2}
({\bf q};s|{\bf t}_{{\rm R},++}(z)|{\bf p};s')$, 
where $E_q,E_p$ and $w_q,w_p$ denote the
relativistic energies of the outgoing/incoming nucleons and pions, 
respectively,  one may conclude, that the spin-nonflip
part $T_1^{\bf n}$ of the remainder amplitude $T_1$ should have the following 
threshold behavior
\eq\label{behaviour}
{\rm Re}\,\biggl\{\,
{\rm e}^{ -2i\alpha\theta_{\rm C}(|{\bf p}|)}\,T_1^{\bf n}\biggr\}=
\frac{B_1}{|{\bf p}|}+B_2\ln\frac{|{\bf p}|}{\mu_c}-8\pi(m_p+M_{\pi^+})\,
{\cal A}+\cdots\, ,
\en
where ${\bf p}$ denotes the relative momentum of $\pi^- p$ pair in the
CM frame, $B_1,~B_2$ do not depend on ${\bf p}$, and
\eq\label{curly-A}
-\frac{2\pi}{\mu_c}{\cal A}=
-d_1+\frac{\mu_0^2 q_0^2}{4\pi^2}\,d_2^2d_3+d_1^2\xi\, 
\en
where ${\cal A}$ is the regular part of $\pi^- p$ scattering amplitude 
at threshold. 

Substituting Eq.~(\ref{curly-A}) into Eq.~(\ref{Re}), we finally arrive at
\eq\label{strong}
\epsilon_{1s}=-2\alpha^3\mu_c^2\, {\cal A}\,(1-2\alpha\mu_c\,
(\ln\alpha-1)\,{\cal A})+\cdots\, ,
\en
where the UV divergence contained in the quantity $\xi$, has been cancelled. 

The equation~(\ref{strong}) is the main result of the present paper - 
it gives the relation between the measured strong energy-level shift of 
the $\pi^- p$ atom and the $\pi^- p$ scattering amplitude at threshold,
defined by Eqs.~(\ref{splitting}) and (\ref{behaviour}). This relation is
valid at $O(\delta)$, and in all orders in the chiral expansion. 
The threshold amplitude still contains the isospin-breaking effects.
Consequently, in order to extract from the experimental data
the $\pi N$ scattering lengths which are defined for 
the case with no isospin symmetry violation, one has to single out the
isospin-breaking effects in the amplitude. To this end, one may invoke ChPT -
below, we present the results of the calculations at the tree level. 

In order to evaluate the $\pi N$ scattering amplitude, we use the effective 
chiral pion-nucleon Lagrangian at $O(p^2)$~\cite{GSS,Meissner,Becher}.
The amplitude is given by the following expression
\eq\label{amplitude}
{\cal A}&=&a_{0+}^++a_{0+}^-+\epsilon=\frac{1}{8\pi(m_p+M_{\pi^+})
F_\pi^2}\,\,\biggl\{
m_pM_{\pi^+}-\frac{g_A^2m_pM_{\pi^+}^2}{m_n+m_p+M_{\pi^+}}
\nonumber\\[2mm]
&+&m_p(-8c_1M_{\pi^0}^2+4(c_2+c_3)M_{\pi^+}^2-4e^2f_1-e^2f_2)\biggr\}\, ,
\en
where $c_i,~f_i$ denote the low-energy constants in the $O(p^2)$ chiral
pion-nucleon Lagrangian, $g_A$ is the nucleon axial-vector coupling
constant~\cite{GSS,Meissner,Becher}, and in our calculations we use the value
$F_\pi=92.4~{\rm MeV}$. The scattering lengths 
$a_{0+}^+,~a_{0+}^-$ are
calculated in the isospin symmetric theory with $e=0$, $m_u=m_d$ and
where, by convention, the masses of the pions and nucleons are taken equal
to the physical masses of the charged pion and proton, respectively.
With this convention, the isospin-breaking correction to the $\pi^-p$ elastic
scattering amplitude at $O(p^2)$ is equal to
\eq\label{epsilon}
\epsilon=\frac{m_p}{8\pi(m_p+M_{\pi^+})F_\pi^2}\,\,\biggl\{
8c_1(M_{\pi^+}^2-M_{\pi^0}^2)-4e^2f_1-e^2f_2\biggr\}\, .
\en

Below, we present the results of the numerical analysis for the
isospin-breaking correction to the $\pi^-p$ atom energy-level shift.
In order to evaluate the second term in the brackets in Eq.~(\ref{strong}),
one may safely approximate ${\cal A}=a_{0+}^++a_{0+}^-$ and use the scattering
lengths given in Ref.~\cite{PSI2}: the total correction coming from this term
amounts up to $+0.66\times 10^{-2}$. 
Further, in order to evaluate the isospin-breaking correction to the
$\pi^-p$ scattering amplitude at threshold given by Eq.~(\ref{epsilon}),
one has to specify the values of the $O(p^2)$ low-energy constants
$c_1$, $f_1$ and $f_2$. We use the value of $c_1$ 
determined from the fit of the elastic $\pi N$ scattering
amplitude at threshold to KA86 data~\cite{private}: 
$c_1=-0.925~{\rm GeV}^{-1}$. The value of the constant $f_2$
can be extracted from the proton-neutron electromagnetic mass
difference~\cite{Reports}: $e^2f_2=(-0.76\pm 0.3)~{\rm MeV}$. The 
determination of the constant $f_1$ from data is however, problematic. For
this reason, in our analysis we have used order-of-magnitude estimate for 
this constant: $-|f_2|\leq f_1\leq |f_2|$.

\begin{figure}[t]
 \vspace{9.1cm}
\includegraphics{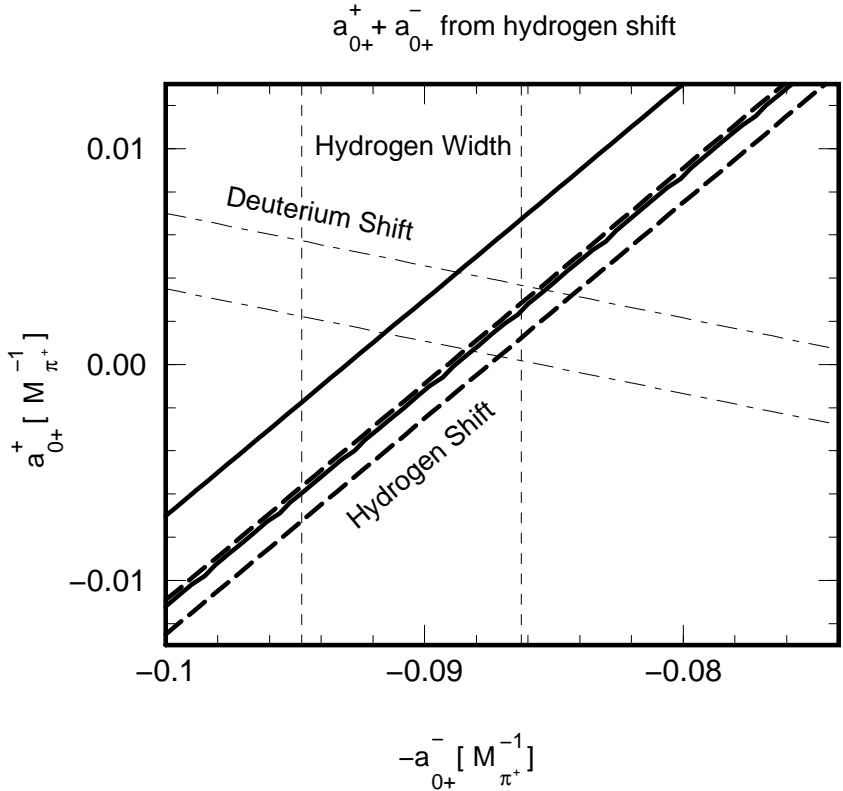}
 \caption{\it
Determination of $\pi N$ scattering lengths from the pionic hydrogen and
pionic deuterium measurements. Solid line corresponds to the energy-level
shift calculations at $O(p^2)$ in ChPT, and the dashed lines - to the
potential model results\cite{PSI1}
    \label{fig1} }
\end{figure}

The prediction for the $\pi^-p$ ground-state energy-level shift based on
$O(p^2)$ calculations in ChPT, strongly deviate from the predictions of the
potential model. The prediction for the total isospin-breaking correction
in the strong energy-level shift, defined according to~\cite{PSI1} 
$\epsilon_{1s}=\epsilon_{1s}^{LO}(1+\delta_{\epsilon})$, where
$\epsilon_{1s}^{LO}$ stands for the leading-order shift given by 
Deser formula~\cite{Deser}, is: $\delta_\epsilon=(-4.8\pm 2.0)\cdot 10^{-2}$
at $O(p^2)$ in ChPT. This result differs more than twice from the 
prediction based on the potential model~\cite{PSI1}:
$\delta_\epsilon=(-2.1\pm 0.5)\cdot 10^{-2}$, and the uncertainty caused
mainly by a poor knowledge of the constant $f_1$, is four times larger
than the estimate of the systematic error in the potential model.
In demonstration of the above discussion,
in Fig.~\ref{fig1} we confront the results of the present analysis with
those of the potential model, using the experimental data on hadronic atoms
taken from Ref.~\cite{PSI2}.

We would like to emphasize that the physical effects encoded in the
low-energy constants $c_1$, $f_1$ and $f_2$, are absent in the potential
model. Namely, $f_1$ and $f_2$ contain the effect of the
direct quark-photon interaction, and the correction proportional to $c_1$ 
contains the effect coming from the dependence of the scattering amplitude on
the quark mass - similar effect in the $\pi^+\pi^-$ case constitutes
the tree-level ``mass-shift'' correction (see,
e.g.~\cite{Sazdjian,Atom,Bern2}). 
For this reason, the discrepancy between the potential model predictions and
the results of the present analysis based on ChPT, does not come to our
surprise. It remains to be seen, how the $O(p^2)$ results in ChPT are 
altered by the loop corrections. A reliable estimate of the constant 
$f_1$ is also desirable.

To conclude, we have derived the general formula that relates the 
energy-level shift of the $\pi^- p$ atom to the $\pi N$ scattering 
amplitude at threshold. Numerical analysis carried out on the basis of this
formula at $O(p^2)$ in ChPT, already indicates at
the necessity to critically reaccess
the values of the $\pi N$ scattering lengths, extracted from the energy-level
shift measurement by means of the potential model-based theoretical
studies.

\vspace*{.5cm}

{\it Acknowledgments.}  
We are grateful to J.~Gasser for the current interest in the work and useful
suggestions. We thank A.~Badertscher, T.~Becher, D.~Gotta, H.-J.~Leisi, 
H.~Leutwyler and U.-G.~Mei\ss ner for useful discussions.
This work was supported in part by the Swiss National Science
Foundation, by TMR, BBW-Contract No. 97.0131 and EC-Contract
No. \, ERBFMRX-CT980169 \, (EURODA$\Phi$NE).

\end{document}